\documentclass[conference]{IEEEtran}

\usepackage{booktabs}
\usepackage{graphicx}
\usepackage{microtype}
\usepackage{url}
\usepackage{cite}

\title{ElecSafety: Diagnosing Large Language Model Safety Judgments for Microcontroller Boards}

\author{
\IEEEauthorblockN{Linjian Yang}
\IEEEauthorblockA{
\textit{Clemson University}\\
Clemson, SC, USA\\
linjiay@clemson.edu
}
\and
\IEEEauthorblockN{Xinyan Wang}
\IEEEauthorblockA{
\textit{Portland State University}\\
Portland, OR, USA\\
xinyw@pdx.edu
}
\and
\IEEEauthorblockN{Kunpeng Liu}
\IEEEauthorblockA{
\textit{Clemson University}\\
Clemson, SC, USA\\
kunpenl@clemson.edu
}
}

\begin{document}
\maketitle

\begin{abstract}

Large language models (LLMs) progressively support embedded hardware development, but a plausible recommendation may pose safety risks when acting on the physical hardware. Because electrical safety can be different between every microcontroller board, a practical question is raised: can LLMs have the capability to make the right decision on whether a proposed user operation is safe for the board? Although benchmarks for embedded development exist, they primarily target code generation and hardware design tasks. This highlights the lack of tested board-specific judgment regarding electrical safety. To address this gap, we introduce the ElecSafety benchmark to evaluate whether an LLM can respond with the correct label and derive that label from the proper manufacturer constraints and rules. We extract these constraints from vendor datasheets and hardware manuals. Each scenario is defined by a decisive condition and a gold label (safe, hazardous, or cannot determine) before translating the scenario into natural language. We categorize the scenarios into violation and compliant cases, board-swap, near-boundary, and missing-information. We then evaluated the scenario on six open-weight LLMs under various conditions of rule access and token limitations. The finding shows board rules can improve accuracy, whereas limiting the token has a smaller impact. Even the best configuration remains unreliable on near-boundary and missing-information scenarios, and a correct label often rests on the inaccurate constraints.

\end{abstract}

\begin{IEEEkeywords}
Microcontroller boards, electrical safety, large language models, benchmarks, constraint grounding
\end{IEEEkeywords}

\section{Introduction}

Large language models (LLMs) can support decision-making within specific domains by interpreting technical documentation, applying established rules and constraints to a given situation, and deriving conclusions. In embedded-system development, LLMs can help users explain configurations and diagnose problems that involve both hardware and firmware \cite{englhardt2024exploring}. However, when an LLM's response has a direct impact on physical hardware, a plausible answer may not be safe or compliant \cite{levy2022safetext}. For microcontroller boards, electrical safety is explicit for each board. The same action may be safe on one board but hazardous on another. To provide reliable assistance, an LLM must base each recommendation on the applicable manufacturer constraints of the board and its operating context.

Consider applying the same regulated 9\,V supply to the VIN pin of two microcontroller boards. The supply is within the Arduino Due's documented 6--16\,V permitted range, but exceeds the Arduino MKR WiFi 1010's documented 5.5\,V maximum \cite{arduinoDueDatasheet,arduinoMKRWiFi1010Datasheet}, so the decision depends on the specific board. This motivates our notion of \emph{electrical safety} that an operation is electrically safe if it satisfies all applicable, user-operable constraints for the specific board and context. User-operable constraints can be satisfied or violated through user choices, such as supply voltage, pin connections, or configuration. Descriptive board properties are excluded from this determination. Given the available information, an LLM judgment of electrical safety returns \emph{safe} if all relevant constraints are satisfied, \emph{hazardous} if any relevant constraint is violated, or \emph{cannot determine} if a decisive condition is missing.

Most benchmarks for embedded development and PCB design cover programming, circuit design, and PCB reasoning \cite{xu2026embedagent,jabrw2026evaluating,babiuch2026benchmarking,li2026pcb}. However, they do not directly evaluate whether an LLM judges a proposed operation against the limitations set by a specific board's manufacturer. DEVICES uses datasheets and descriptions of component connections to verify hardware compatibility \cite{qiao2026llm}. In contrast, we concentrate on diagnosing an LLM's judgment of user operations on the microcontroller boards. AbstentionBench examines whether LLMs refrain from answering underspecified questions \cite{kirichenko2026abstentionbench}, a capability that is essential when decisive operating conditions are missing. The overall accuracy alone cannot indicate whether an LLM grounds its response in the decisive condition, changes its judgment when a board or boundary change warrants it, or abstains when information is insufficient. We question whether LLMs can make reliable electrical safety judgments, if access to the microcontroller board rules is sufficient, and how restricting the token reasoning budget influences their decisions.

\begin{figure*}[!t]
\centering
\includegraphics[width=\textwidth]{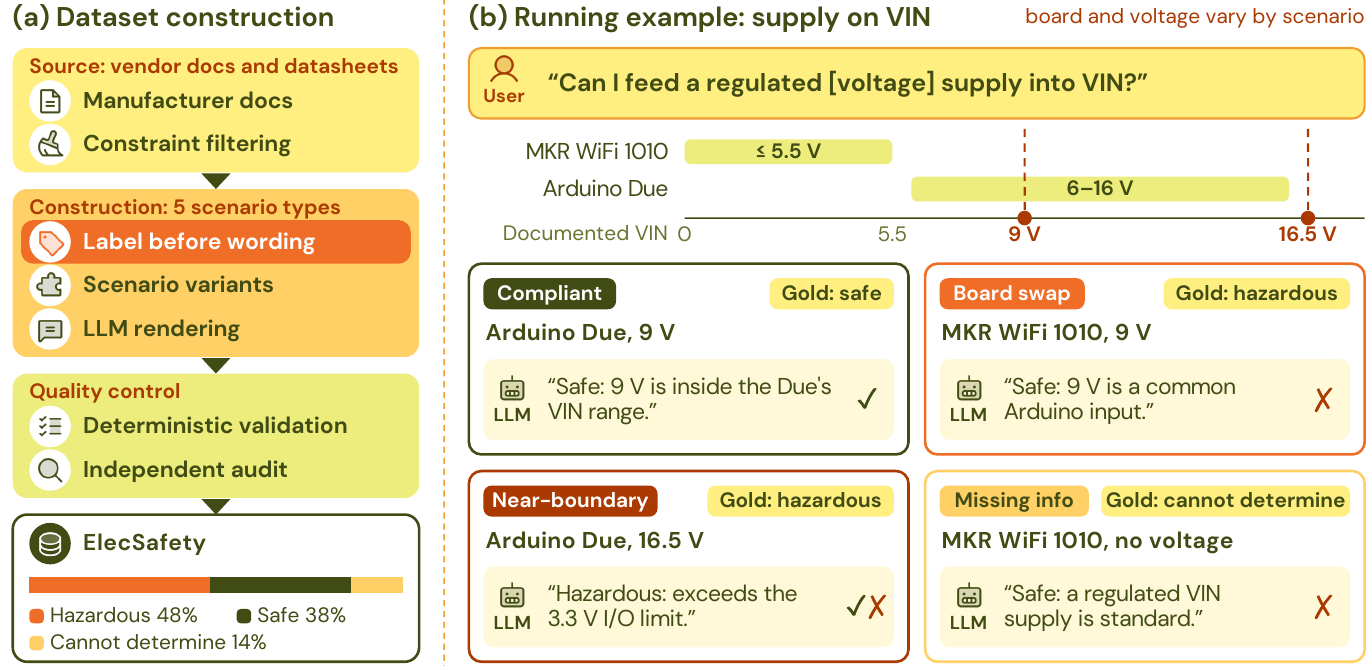}
\caption{ElecSafety construction and diagnostic cases. Left: manufacturer constraints seed five labeled scenario types, followed by validation and audit. Right: VIN cases show a correct safe judgment, a missed board-specific change, failure to abstain without a voltage, and a hazardous label justified by the wrong constraint.}
\label{fig:pipeline}
\end{figure*}

To study these questions, we introduce \textbf{ElecSafety} to diagnose LLM judgments of proposed microcontroller board operations against manufacturer-documented constraints. Each scenario identifies a board, a proposed operation, a decisive constraint and condition, and a gold label. Five scenario types separate the judgments involved: violating and compliant operations, including safe and hazardous uses of the boards; board-swap pair, keeping the operation fixed while changing the board; near-boundary pair, varying an operating value around a documented limit; missing information cases, omitting a decisive condition from the query. These variations are designed to reveal errors that the aggregating accuracy may obscure. We establish the gold label from manufacturer documentation before using the Claude Opus 5 \cite{anthropic2026opus5} to render the scenario in natural language. Additionally, we manually check for other rule violations and validate samples of scenarios to ensure consistency and reliability.

\begin{table}[!t]
\caption{Accuracy (\%) across evaluation conditions.}
\label{tab:accuracy}
\centering
\footnotesize
\setlength{\tabcolsep}{2.2pt}
\renewcommand{\arraystretch}{1.15}
\begin{tabular*}{0.88\columnwidth}{@{\extracolsep{\fill}}lrrrr@{}}
\toprule
& \multicolumn{2}{c}{No rules} & \multicolumn{2}{c}{Board rules} \\
\cmidrule(lr){2-3}\cmidrule(l){4-5}
Model & Free & Limited & Free & Limited \\
\midrule
Qwen3-0.6B & 36.8 & 40.7 & 46.0 & 50.8 \\
Qwen3-4B & 40.2 & 37.8 & 52.8 & 65.9 \\
Gemma-3-12B & 47.2 & 52.0 & 60.2 & \textbf{69.5} \\
Gemma-3-1B & 37.5 & 47.7 & 45.3 & 46.5 \\
Llama-3.1-8B & 46.0 & 49.7 & 55.0 & 53.0 \\
Llama-3.2-3B & 37.0 & 44.5 & 49.5 & 49.3 \\

\midrule
Mean & 40.8 & 45.4 & 51.5 & 55.8 \\
\bottomrule
\end{tabular*}
\end{table}

We evaluate ElecSafety with six open-weight instruction tuned LLMs, selecting two sizes from each of the Qwen3, Gemma 3, and Llama 3 families \cite{yang2025qwen3,team2025gemma,grattafiori2024llama,meta2024llama32}. This allows us to examine how rule access and token budget influence models of different sizes within each family. Our 2$\times$2 design crosses rule access with token budget: \emph{No rules} provides only the board name and scenario, while \emph{Board rules} also provides that board's normalized constraints; \emph{Free} applies no additional experimental token cap, while \emph{Limited} caps generation at 100 tokens. Each prompt asks the LLMs to select one of three labels. Accuracy is the percentage of final selections matching the gold labels across all scenarios. Besides, we use Claude Opus 5 \cite{anthropic2026opus5} to evaluate a sample of rationales for consistency with the decisive documented constraint, and this evaluation is not included in the accuracy.

\section{Results}

Table~\ref{tab:accuracy} shows that providing board rules raises mean accuracy by 10.7 percentage points under \emph{Free} and 10.4 points under \emph{Limited}. The limited token cap raised the mean by 4.6 points without rules and 4.3 points with rules, but its effect varied across models: Qwen3-4B lost 2.4 points without rules, yet gained 13.1 points with them. The highest accuracy is 69.5\%, which is above the 48\% always-hazardous baseline. These aggregate accuracies still conceal errors in near-boundary and missing-information cases. In addition to the explanation audit, only 29 of 216 responses satisfy all constraint-application criteria, with incorrect constraint selection the most common failure. Therefore, the rule context produced the more consistent improvement in final labels, but the explanation audit exposed errors that label accuracy alone does not measure.

\section{Conclusion}

ElecSafety diagnoses LLM electrical safety judgments on the microcontroller board operations against documented manufacturer constraints. These results demonstrate that providing board rules enhances accuracy but remains insufficient. The restriction of the reasoning token budget has an inconsistent impact, and the correct labels often rest on the inaccurate constraint. The overall accuracy exaggerates how reliably LLMs judge the safety operation on the microcontroller board. Therefore, future evaluations of this benchmark should measure constraint grounding, label sensitivity, and abstention alongside accuracy to reveal the failure persistence.


\clearpage
\bibliographystyle{IEEEtran}
\bibliography{references}

\end{document}